\documentclass[a4paper,11pt]{article}
\pdfoutput=1 

\usepackage{jinstpub} 
\usepackage{amsmath}
\usepackage{float}
\title{Effects of Temperature Anneal Cycling on a Cryogenically Proton Irradiated CCD}

\author[a,1]{S. Parsons, \note{Corresponding author.}}
\author[a]{T. Buggey}
\author[a]{A. Holland}
\author[c]{S. Sembay}
\author[a]{G. Randall}
\author[a]{O. Hetherington}
\author[d]{D. Yeoman}
\author[a]{D. Hall}
\author[b]{P. Verhoeve}
\author[a]{M. Soman}

\affiliation[a]{Open University, Centre of Electronic Imaging,\\Milton Keynes, UK}
\affiliation[b]{Directorate of Science of the European Space Agency, ESTEC,\\Noordwijk, The Netherlands}
\affiliation[c]{University of Leicester, \\Leicester, UK}
\affiliation[d]{Teledyne-e2v, \\Chelmsford, UK}

\emailAdd{Steven.Parsons@open.ac.uk}

\abstract{
	Throughout a typical Earth orbit a satellite is constantly bombarded by radiation with trapped and solar protons being of particular concern as they gradually damage the focal plane devices throughout the mission and degrade their performance. To understand the impact the damage has on CCDs and how it varies with their thermal history a proton radiation campaign has been carried out using a CCD280. The CCD is irradiated at 153~K and gradually warmed to 188~K in 5~K increments with Fe$^{55}$ X-ray, dark current and trap pumping images taken at 153~K after each anneal step. 
	
	The results show that despite the trap landscape changing throughout the anneal it has little impact on parallel charge transfer inefficiency. This is thought to be because most traps are unaffected and a lot of those that do anneal only move from the continuum between distinct trap species and into a nearby divacancy trap “peak” whose emission time constant is similar enough to still impact the CTI.  
	
	In terms of using a CCD280 or similar devices in a mission the CTI being unaffected by thermal annealing up to 188~K means that any CTI correction needed as the radiation damage builds up does not have to take into account the thermal history of the focal plane. However, it is possible that a significant amount of annealing will occur at temperatures greater than 188~K and care should be taken when a mission is operating in this range to gather accurate pre-flight data
}

\keywords{CCD, radiation damage, temperature annealing, radiation damage, X-ray detectors, trap pumping, charge transfer inefficiency}

\begin{document}
\maketitle
\flushbottom

\section{Introduction}
\label{sec:intro}
Charge-Coupled Devices (CCDs) are a well-established focal plane detector of choice and have been flown in a large variety of space missions due to their ability to provide high quality data whilst maintaining low noise and dark current (DC). However, the space radiation environment they operate within causes the silicon of the detector to become damaged over time. This is mainly caused in missions using Earth orbits such as SMILE ~\cite{RaabSMILE} by high energy protons which disrupt the uniform lattice structure by knocking a silicon atom out and creating a “vacancy”. The vacancy migrates through the lattice until it can form a stable defect such as the divacancy or Silicon E-centre, this then creates additional energy levels or “traps” between the conduction and valence bands. These traps can capture electrons from signals as they are transferred through the device and then release them later depending on the individual trap’s emission time constant, the Shockley-Read-Hall theory explains the capture and release mechanics of traps in detail ~\cite{ShockleyRecomb} ~\cite{HallRecomb}. It is this effect that increases the charge transfer inefficiency (CTI) which causes degradation of the images through charge loss from slower traps or smearing from faster ones. 

The emission time constant of a specific trap varies with temperature, therefore, to accurately predict the impact of traps present in an irradiated CCD throughout its mission the intended operational temperature conditions must be used for the characterisation. Traditionally this has been achieved for space missions by irradiating at room temperature and then cooling the device down under vacuum and testing. However as shown in ~\cite{Moll} a device will start to anneal almost immediately after irradiation due to the activation energy of the traps being met by thermal energy from the environment causing the trap landscape (and hence device performance) to change irreversibly. Owing to this it has become more commonplace recently to irradiate a device cold, often at the operating temperature of the instrument and maintain that temperature throughout testing to keep the mission-like landscape and therefore improve the accuracy of the results.
 
The work presented in this paper has been carried out by the Centre for Electronic Imaging at The Open University to investigate the performance of a CCD280 after being irradiated by 7.4~MeV protons at the nominal SMILE operating temperature of 153~K to a 10~MeV equivalent proton fluence of 2.82×10$^{9}$~pcm$^{-2}$ as part of the SMILE radiation damage studies. The proton fluence is approximately 70~\% of the end of life (EOL) fluence the SMILE SXI CCDs will receive and is calculated through Spenvis simulations using the predicted orbit and solar conditions (at time of irradiation). This paper focuses on the impact of isothermal annealing on the DC, trap landscape and the parallel CTI by carrying out DC, trap pumping and X-ray CTI measurements over 153 to 188~K which covers, with margins, the approximate range the SMILE focal plane will vary across throughout an orbit.

\section{Methodology}
\subsection{Device Under Test}
The back illuminated 1552 x 1280 pixel CCD280 manufactured by Teledyne-e2v and used throughout this campaign is an example of the test vehicle for the PLATO mission and is functionally very similar to the larger 4510 x 4510 pixel CCD270 flight devices, which have been cut from the same wafer to provide uniformity, the specifications of the CCD280 are summarised in Table ~\ref{Tab:CCD280Specs} ~\cite{EndicottPLATO}.

\begin{table}[htbp]
	\centering
	\caption{\label{Tab:CCD280Specs} CCD280 design specifications.}
	\smallskip
	\begin{tabular}{|c|c|}
		\hline
		Specification & Value\\
		\hline
		Manufacturing Company & Teledyne-e2v \\
		Silicon Substrate & SiC\\
		Image area size & 776 rows x 1280 columns \\
		Store area size & 776 rows x 1280 columns \\
		Pixel size & 18 x 18 $\mu$m \\
		Sensitive Silicon Thickness & 16 $\mu$m \\
		Serial registers & 1 \\
		Line Transfer Time & 67 $\mu$s\\
		Noise (4 MHz) & 20 e $^{-}$ rms\\
		Dark Signal (203 K) & 0.4 e$^{-}$.pix$^{-1}$.s$^{-1}$\\
		Full Well Capacity & 900 ke$^{-}$ \\

		\hline
	\end{tabular}
\end{table}

The PLATO CCDs have served as a precursor to the SMILE devices which have subsequently been optimised to maximize their efficiency for the low flux, low energy signals the SXI will detect. 

The CCD280 has a store area which covers half of the device and can be run in either full frame or frame transfer mode. The results presented in this paper are from a CCD280 run in native pixel full frame mode. 
    
\subsection{Irradiation Schedule}

\begin{figure}[htbp]
	\centering 
	\includegraphics[width=\textwidth,trim=1 0 0 0,clip]{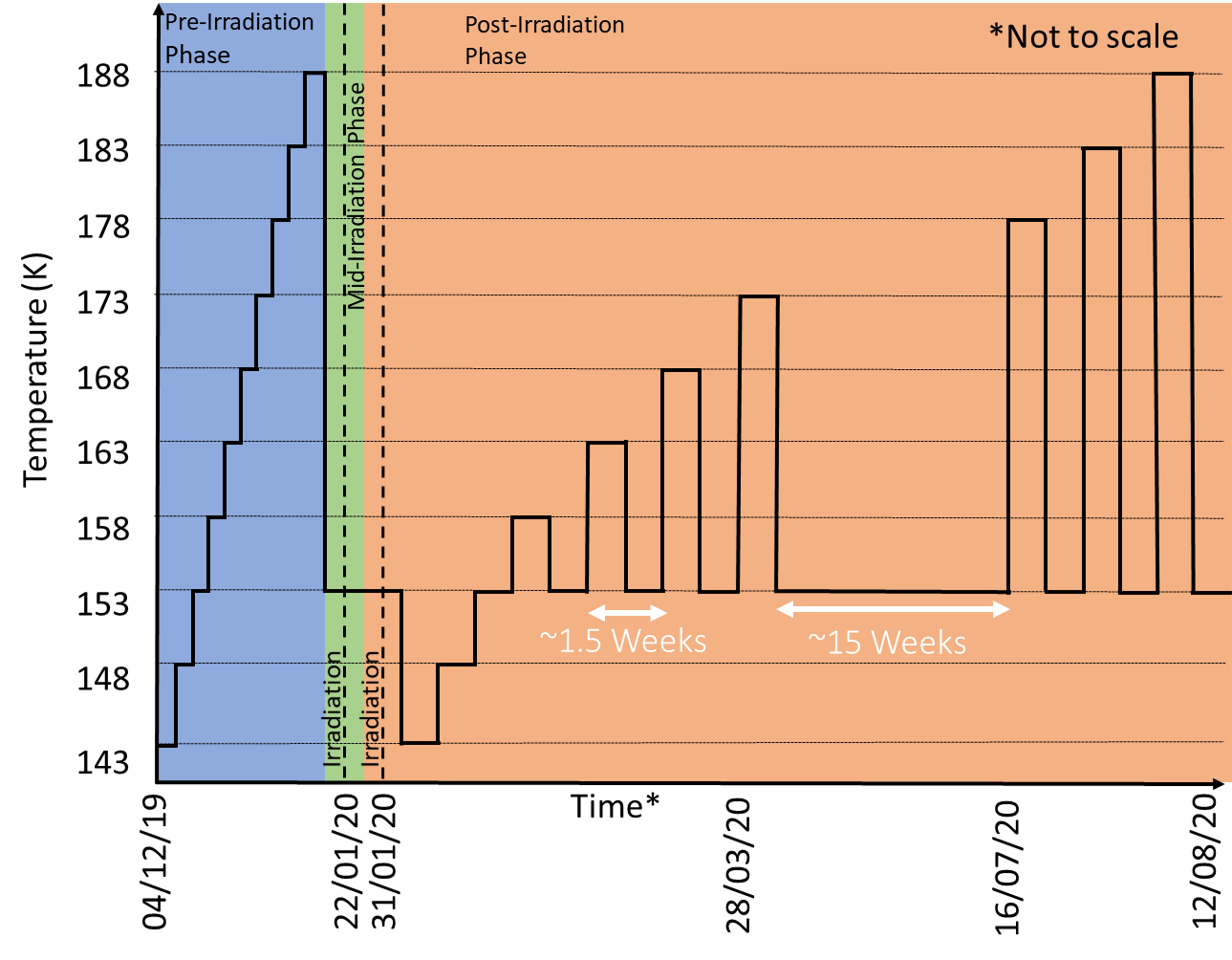}
	\caption{\label{fig:Schedule} Radiation damage campaign characterisation schedule. Measurements are taken at each horizontal section. }
\end{figure}

Prior to irradiation the CCD goes through a characterization phase where the dark current, bright defects, charge injection uniformity, trap landscape, and CTI are measured across the 143 to 188~K temperature range. The device is then irradiated in two steps up to the 70~\% EOL fluence predicted from the modelling of the radiation environment for the SMILE mission. After the first irradiation the device is characterized at the nominal operating temperature and then irradiated a second time. A characterization stage is indicated in Figure ~\ref{fig:Schedule} at each horizontal section and it shows that after irradiating the CCD temperature is lowered to 143~K and characterized before gradually increasing up to 188~K in 5~K steps. After each warming stage above the irradiation temperature the device is re-characterized at 153~K to assess the impact of any annealing that has taken place. Each stage lasts for approximately 5 days before the temperature changes, after the 173~K warming stage the CCD is held at 153~K for 15 weeks with a characterization before and after. 

\subsection{Experimental Setup}
\begin{figure}[htbp]
	\centering 
	\includegraphics[width=\textwidth,trim=1 0 0 0,clip]{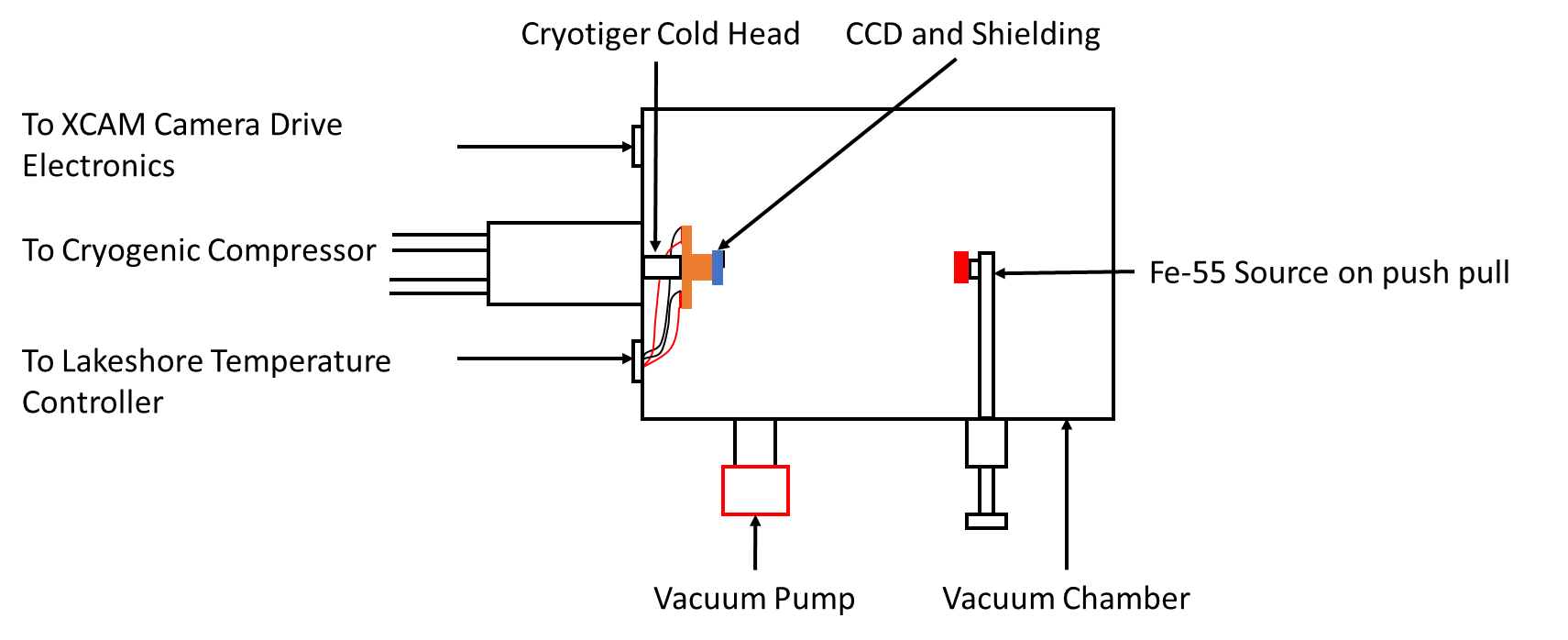}
	\caption{\label{fig:SetupSchematic} Schematic of vacuum chamber design showing cooling systems and Fe$^{55}$ X-ray source used at the test facility.}
\end{figure}

\begin{figure}[H]
	\centering 
	\includegraphics[width=\textwidth,trim=1 0 0 0,clip]{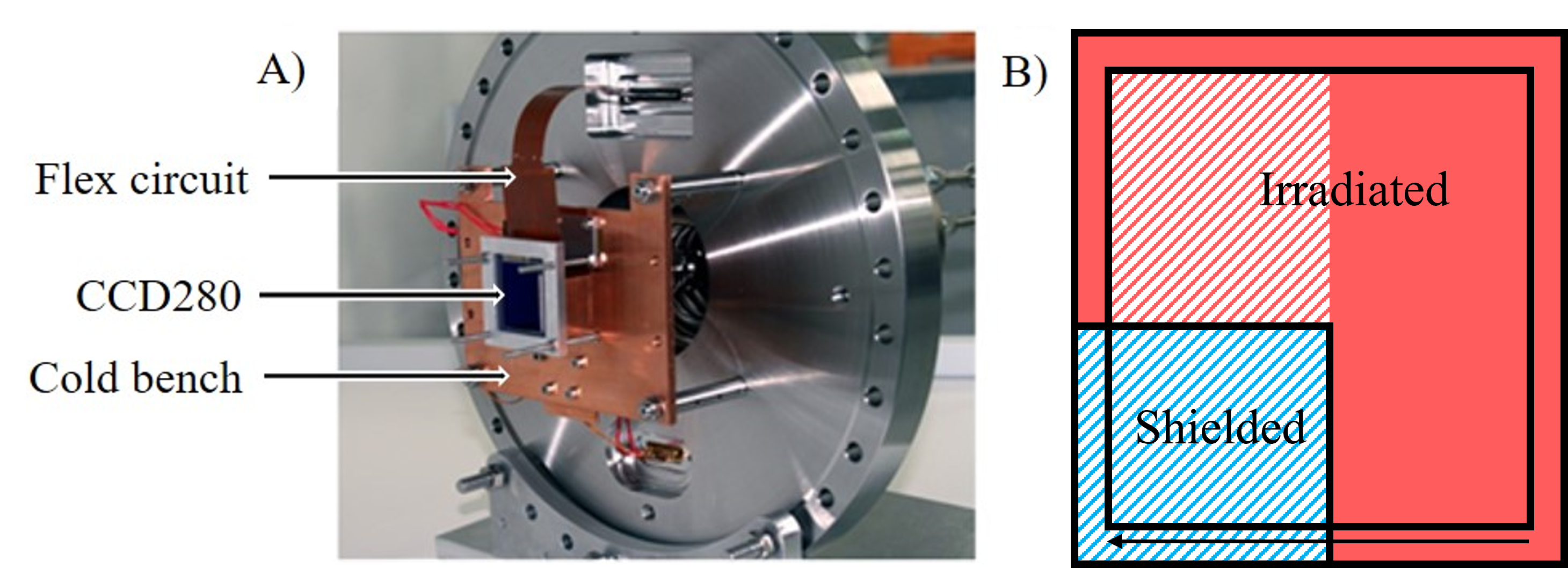}
	\caption{\label{fig:Coldhead} CCD280 A) Mounted onto copper cold bench prior to loading into vacuum chamber B) Shielding layout during proton irradiation, hatched blue is shielded area and hatched red is area of device used to measure pCTI.}
\end{figure}

The equipment setup used is shown in Figure ~\ref{fig:SetupSchematic}, where the device is mounted inside the vacuum chamber by clamping the CCD onto a copper cold bench which is connected to a CryoTiger® refrigeration system (PT-30), as in Figure ~\ref{fig:Coldhead}a. The thermal control is provided through using a resistive heater in thermal contact with the cold bench, allowing an operating temperature range of 138 to 193 K before overloading the cooling capacity of the CryoTiger®. Temperature control is provided to within 0.1 K using a feedback system, composed of a Lakeshore 325 temperature controller, the heater, and a 1 k$\Omega$ platinum resistance thermometer in the CCD package. A removeable Fe$^{55}$ source is used to provide a known signal (5898 eV) for calibration and CTI measurements. Clocking and biasing are provided by an XCAM Ltd USB2REM1 camera drive box in conjunction with custom drive software designed to operate using MATLAB.

The system relies on shields which are placed in front of the CCD with a 2 mm gap between the detector surface and the shield. The region to be shielded in the CCD280 is shown in Figure ~\ref{fig:Coldhead}b as hatched blue and uses a 1 cm thick steel shield to protect the image area around the readout electronics and part of the serial register, this enables the post irradiation parallel CTI to be measured in the hatched red area with minimal serial CTI contribution. During the irradiation and subsequent testing, the system is powered continuously and has been designed to have sufficient thermal mass of copper and steel to limit the impact of any short-term power loss on the temperature of the device.


\subsection{Test Facility}

The cyclotron facility at Synergy Health, Harwell, England is used for the irradiation campaign, the beam-line and device setup are summarised in Table ~\ref{Tab:IrradSpecs}. A total fluence of  2.82x10$^{9}$ protons (10 MeV equiv.).cm$^{-2}$ is provided through one irradiation of 2.06x10$^{9}$ proton (10 MeV equiv.).cm$^{-2}$ and one of 0.76x10$^{9}$ proton (10 MeV equiv.).cm$^{-2}$ with the device kept at 153~K and biased throughout the campaign.

\begin{table}[htbp]
	\centering
	\caption{\label{Tab:IrradSpecs} CCD280 proton irradiation characteristics.}
	\smallskip
	\begin{tabular}{|c|c|c|}
		\hline
		Device No.                                      & \multicolumn{2}{|c|}{280 (14512-21-02)} \\
		\hline
		Irradiation step                                & 1                 & 2                 \\
		Device temperature during irradiation (K)       & 153               & 153               \\
		CCD Biasing                                     & Running           & Running           \\
		Proton Energy (MeV)                             & 7.4               & 7.4               \\
		10 MeV proton equivalent fluence (protons.cm$^{-2}$) & 2.06×10$^{9}$          & 0.76×10$^{9}$          \\
		Flux (Protons.cm$^{-2}$.s$^{-1}$)                         & 1.84x10$^{7}$          & 1.84x10$^{7}$ \\   
		\hline     
	\end{tabular}
\end{table}

\subsection{Test Conditions}
\subsubsection{X-ray CTI}

After an integration period the device is readout, this involves a parallel transfer which moves all the charge clouds one row closer the the readout register and 1280 serial transfers as the pixels are moved into the readout electronics, this is repeated for the 1552 rows. Parallel CTI refers to the charge that is lost whilst moving down through the rows and serial CTI to the charge lost whilst transferring across the columns.

\begin{figure}[htbp]
	\centering 
	\includegraphics[width=\textwidth,trim=1 0 0 0,clip]{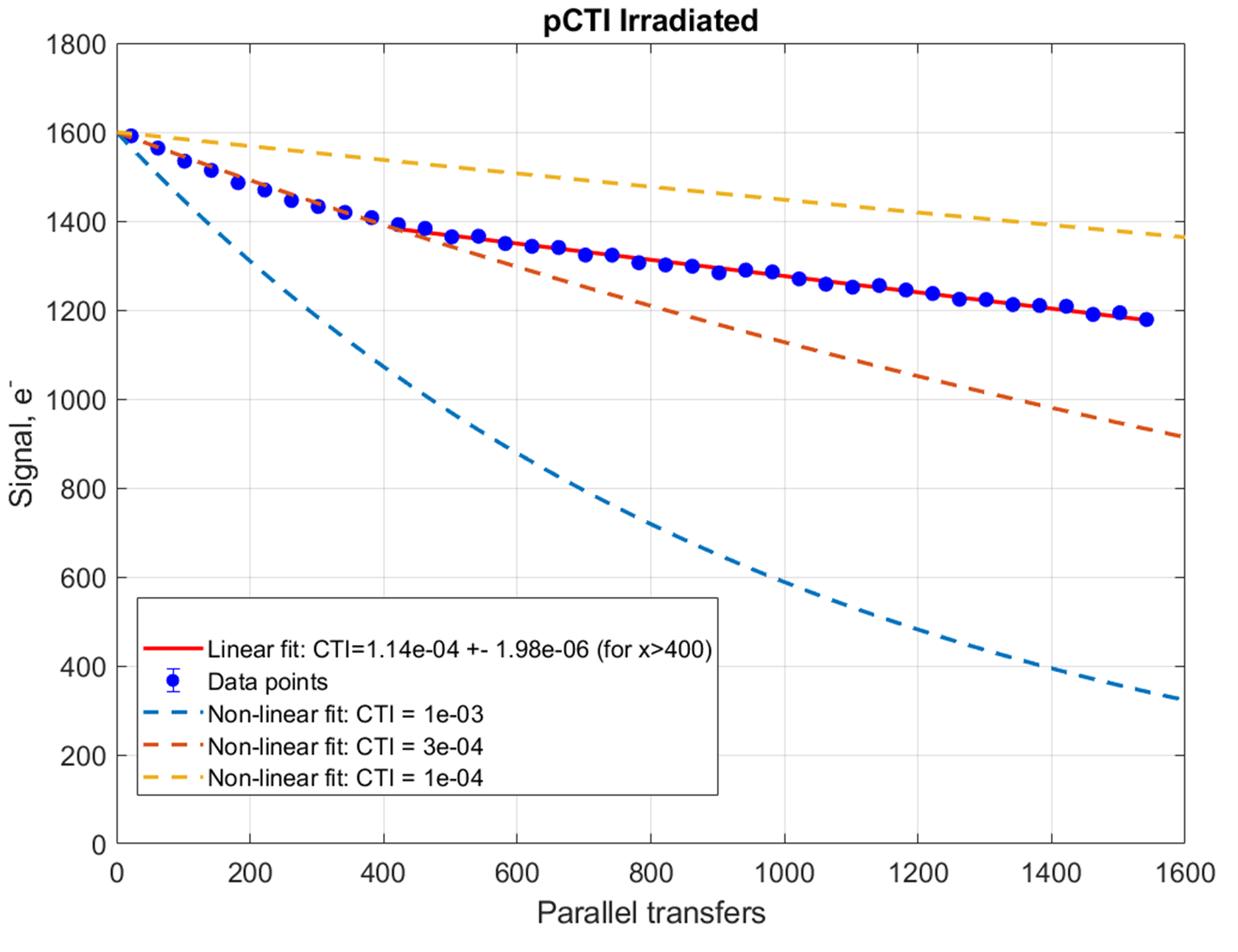}
	\caption{\label{fig:CTIfit} 173 K X-ray parallel CTI with overlaid linear and non-linear fits showing how the inclusion of the first $\approx{400}$ parallel transfers can generate poor fits when slow traps are present.}
\end{figure}

The focus of this paper is on how the parallel CTI responds to temperature annealing and it has been measured using stack line trace methods with Fe$^{55}$ X-ray photons. The test conditions used to capture these images are summarized in Table ~\ref{Tab:CCD280TestCond}.
 To analyse the images, all X-rays are identified, and then the X-rays where signal is split across multiple pixels from charge diffusion are discarded. A further down-selection process is used to select the single-pixel X-rays that have an energy similar to the characteristic Fe$^{55}$ k$\alpha$ spectral line. X-rays are binned by row location and the photo-peak position measured through a Gaussian fit which is then combined with the position in the image to generate Figure ~\ref{fig:CTIfit} which shows how the measured X-ray signal decreases the further a charge packet has to travel through the device, the error on the CTI is calculated from combining the Gaussian fit error and the line fit. 

\begin{table}[htbp]
	\centering
	\caption{\label{Tab:CCD280TestCond} X-ray CTI test conditions used throughout the cryogenic CCD280 proton irradiation campaign.}
	\smallskip
	\begin{tabular}{|c|c|}
		\hline
		Variable & Setting\\
		\hline
		Integration Time (s) & 300 \\
		X-ray density (X-rays.pix$^{-1}$) & 1/2500 \\
		Temperature (K) & 153 to 188 \\
		Readout & Full frame\\
		Number of Images per variable & 50 \\
		Charge Injection Signal level (e$^{-}$) & $\approx{8500}$ \\
		\hline
	\end{tabular}
\end{table} 



The CTI is calculated by fitting to this plot, ideally using the non-linear Equations ~\ref{eq:CTE} and ~\ref{eq:CTI} .

\begin{equation}
\label{eq:CTE}
Charge Remaining = signal * CTE ^n \
\\
\end{equation}
\textit{CTE $ = $ Charge Transfer Efficiency, n $ = $ number of transfers}
\\

\begin{equation}
\label{eq:CTI}
CTI = 1 - CTE \
\\
\end{equation}
\\

However as shown by the dashed overlaid traces in Figure ~\ref{fig:CTIfit} a good fit is not possible at higher values of n which indicates that another mechanism is affecting the CTI which is not accounted for in Equation ~\ref{eq:CTI}. The same initial non-linear feature for low numbers of transfers has been seen previously in testing for CCDs on the Euclid VIS instrument at longer integration times ~\cite{Silicondefects}. This effect is thought to be caused by the signal electrons in that region seeing a lower number of longer-emission-time (slow) traps due to travelling through fewer unfilled slow traps; in general, the X-ray signal will travel through the number of slow traps between X-ray events (as once filled they can no longer trap charge). As a fit using all the parallel transfers will give an inaccurate measurement, the CTI in this study is therefore calculated with the first 400 rows nearest the readout excluded. This means that the X-ray CTI values quoted only assess the charge lost from the fast traps due to the fit ignoring the initial step change and the actual change will vary with the integration time. Whether a trap is fast or slow depends on the relative magnitude of the parallel transfer time and the trap emission time. Traps which emit between line transfers (emission time constants < 67~$\mu$s) are considered fast in this case, with trap emission time constants > 67~$\mu$s considered slower. Trap emission time constants do have a spread however (across each individual species and even more so between separate species), so caution must be taken when creating distinct separations between those with "fast" and "slow" emission time constants, and the affect that they have on CTI.


The CCD280 incorporates a charge injection structure at the top of the image array that can be used to inject a fixed level of charge into the row at the top of the image area (furthest from serial register) to be clocked through the pixel array with the aim of filling any empty traps and hence improve the CTI.  In the images testing the impact of charge injection, charge is injected and transferred through the array to the row before the readout register and is held there throughout the integration time before being readout as the first row in the image.


\subsubsection{Trap Pumping}

The trap pumping technique ~\cite{PumpingTrapProperties} ~\cite{PumpingPchannel} ~\cite{BushEMCCD} ~\cite{HallSiliconDI} ~\cite{Woodpchannel} ~\cite{SkottfeltEuclid} ~\cite{Mostek} allows the emission time constants of the intrinsic and radiation-induced defects within the silicon CCD to be calculated, which can then be collated to provide an overall representation of the spread of emission time constants within the device. The underlying theory is based upon Shockley-Read-Hall recombination statistics ~\cite{ShockleyRecomb}~\cite{HallRecomb}.

\begin{figure}[htbp]
	\centering 
	\includegraphics[width=\textwidth,trim=1 0 0 0,clip]{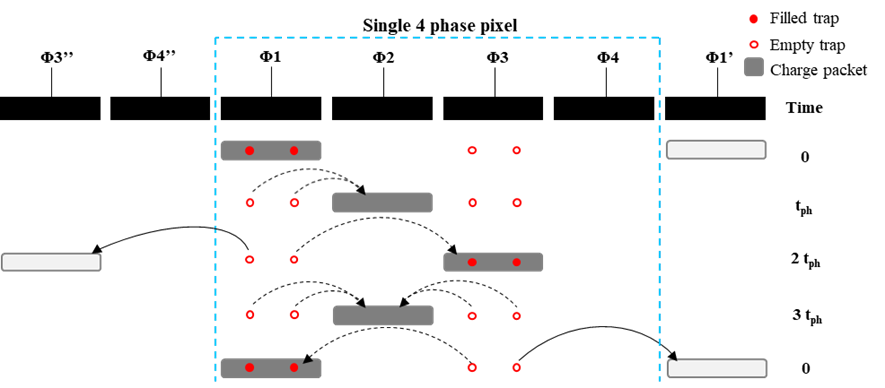}
	\caption{\label{fig:Pumpingseq} Defect capture and emission dynamics for a single trap pumping cycle as part of the 1-2-3-2-1 trap pumping scheme. Note that dashed and solid black arrows represent emission into the source and non-source charge packet, respectively.}
\end{figure}

The trap pumping technique relies upon the rapid, sequential movement of charge through a designated area of the device. A known number of electrons are added to a selected number of pixels in the device, often via the charge injection method. Once added, different trap pumping schemes are used to probe different areas of each pixel. Figure ~\ref{fig:Pumpingseq} shows the 1-2-3-2-1 trap pumping scheme, with charge initially stored under phase $\phi$1. Defects under this phase initially capture electrons, as seen by the filled red circles in Figure 3. After a time t$ _{ph} $, charge is moved to phase $\phi$2 and stored under this phase. During charge transfer and storage, the filled defect can release its electron at any time, with the released electron preferentially joining the closest charge packet (where the highest potential is also located). In this case, the closest charge packet and highest potential is still the source charge packet, so there is no net movement of charge (indicated by the dashed arrows). Whether the defect is in the left- or right-hand side of phase $\phi$1 is irrelevant, both defect locations are still closest to the source charge packet and hence there is still no net movement of charge. The pumping sequence continues, with charge now moved into phase $\phi$3. Defects under phase $\phi$1 can still be filled by stored electrons (depending on their emission time constant) and will still emit into the closest charge packet. For the case of defects on the right-hand side of $\phi$1, the closest charge packet is still the source charge packet, so once again no net movement of charge is seen. However, defects located in the left-hand side of phase $\phi$1 are now closest to the charge packet in the adjacent pixel, with emitted electrons now joining this non-source charge packet (indicated by the solid black line in Figure ~\ref{fig:Pumpingseq}). Overall, this means that an electron has been moved, or “pumped”, from the pixel enclosed by the blue dashed line, into an adjacent pixel, from a single pump cycle.

\begin{figure}[htbp]
	\centering 
	\includegraphics[width=\textwidth,trim=1 0 0 0,clip]{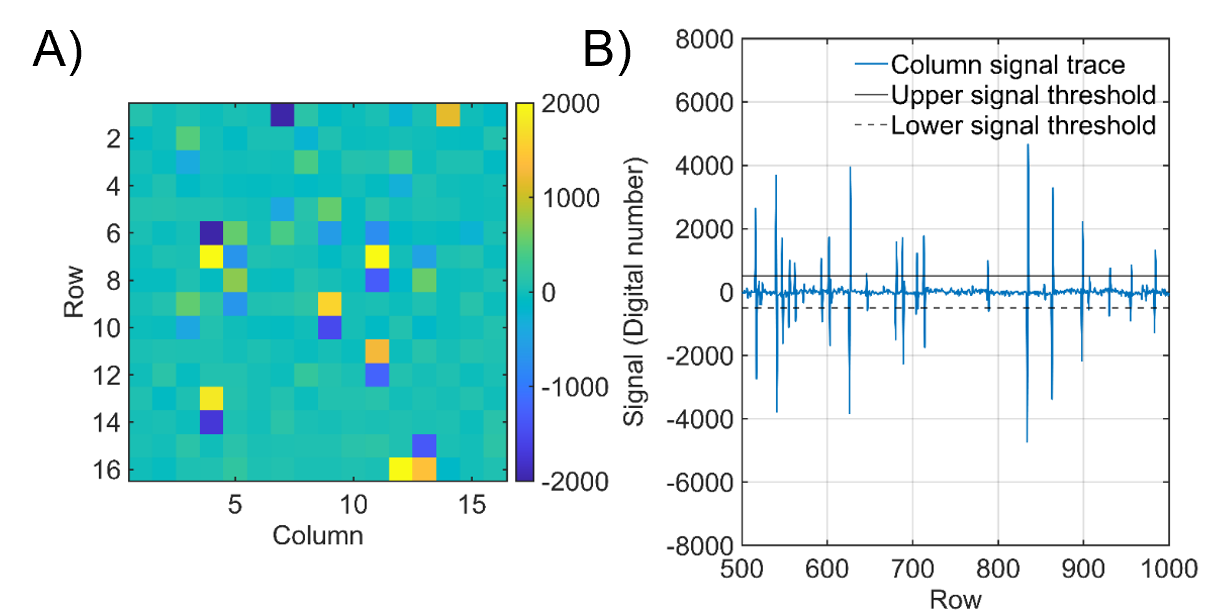}
	\caption{\label{fig:Dipoles} (A) Charge dipoles that are visible after a background signal subtraction. Note that the orientation of the dipoles is not consistent, suggesting that some defects are in different sub-phase locations within each pixel. (B) Changing pixel signal as a function of row number across a single image column, with the upper and lower limit for dipole selection shown.}
\end{figure}

\begin{figure}[htbp]
	\centering 
	\includegraphics[width=\textwidth,trim=1 0 0 0,clip]{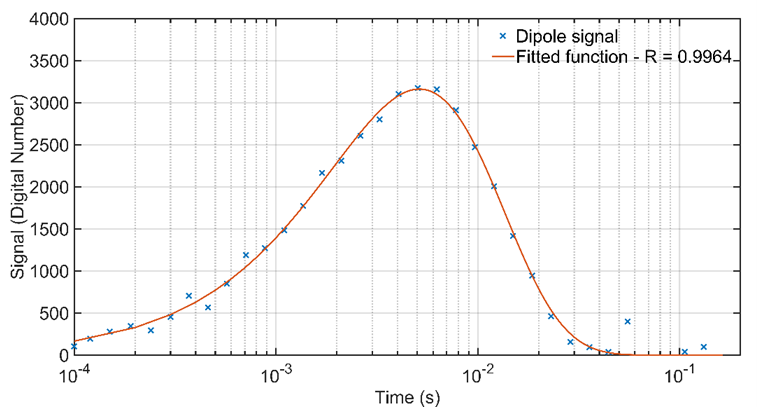}
	\caption{\label{fig:DipoleIntensity} Dipole intensity curve from a single defect measured at a range of pumping times (blue crosses). Overlaid in red is the fitted function, showing an excellent match to the experimental data points.}
\end{figure}

When the trap pumping schemes are implemented, utilizing up to 10,000 pump cycles, distinct patterns of signal are seen within each trap pumped image, shown in Figure ~\ref{fig:Dipoles}a. Each trap probed is revealed by two adjacent pixels (in the column direction) showing a contrast of high and low signal, named charge dipoles. By analysing each column in each image (Figure ~\ref{fig:Dipoles}b) it is possible to locate each trap and proceed with the next stage of analysis. For each defect, the signal of the charge dipole is plotted as a function of the parallel pumping delay (t$ _{ph} $), and shows a characteristic curve as seen in Figure ~\ref{fig:DipoleIntensity}.

To extract the emission time constant, the pumping efficiency of the defects (shown in Figure ~\ref{fig:DipoleIntensity}) must be described mathematically. The probability that a defect will pump (P$ _{p} $) between the times of t$ _{1} $ = t$ _{ph} $ and t$ _{2} $ = 2t$ _{ph} $ (as shown in Figure ~\ref{fig:Pumpingseq}) is ~\cite{PumpingTrapProperties}:

\begin{equation}
\label{eq:PumpProb}
P_{p} = \exp (\frac{-t_{ph}}{\tau_{e}}) -\exp (\frac{-2t_{ph}}{\tau_{e}})  \
\\
\end{equation}

where $\tau_{e}$ is the emission time constant of the defect which is pumping. Note that Equation ~\ref{eq:PumpProb}  is specific to a four-phase pumping scheme (which is applicable to a four-phase CCD like the CCD280), specifically 1-2-3-2-1, and the probability of pumping equation will change if a different trap pumping scheme is used. After several pumping cycles, the quantity of charge pumped, or intensity (I) of the charge dipole is given by the equation ~\cite{PumpingTrapProperties}:

\begin{equation}
\label{eq:DiIntensity}
I = NP_{p} = NP_{c}( \exp (\frac{-t_{ph}}{\tau_{e}}) -\exp (\frac{-2t_{ph}}{\tau_{e}})) \
\\
\end{equation}

where N = number of pump cycles, and Pc = probability of capture of the defect. By fitting Equation ~\ref{eq:DiIntensity} to the dipole intensity curve in Figure ~\ref{fig:DipoleIntensity} (red curve), both the probability of emission and capture can be extracted for that single defect. Mathematically, the peak of the curve in Figure ~\ref{fig:DipoleIntensity} can be described by the differential:

\begin{equation}
\label{eq:CurvePeak}
\frac{dI}{dt} = 0 \
\\
\end{equation}

Substitution of Equation ~\ref{eq:CurvePeak} into Equation ~\ref{eq:DiIntensity}, with subsequent rearranging, leads to:

\begin{equation}
\label{eq:EmissionTime}
\tau_{e} = \frac{-t_{ph(max)}}{\ln(2)} \
\\
\end{equation}

Equation ~\ref{eq:EmissionTime} provides a means of choosing a suitable range of pumping times (t$ _{ph} $) that will probe the emission time constant of a specific defect, to attain a curve like the one shown in Figure ~\ref{fig:DipoleIntensity}. In this study, Equation ~\ref{eq:EmissionTime} was used to choose a suitable range pumping times that focused on the double negative state of the divacancy as well as emission time constant ranges between the divacancy and the unknown defect. Probing longer emission time constants than the unknown would be ideal, as defects with slower emission time constants also affect parallel CTI. However, time constraints on the overall testing process naturally placed a limit on the amount of time assigned to trap pumping activities.

Once the emission time constant of each defect is extracted, the fitted defects can be binned into a histogram. The histogram, often called the “trap landscape”, provides a precise representation of the quantity and distribution of defects within the emission time constant range probed. The trap landscape can then be used to verify key trends in CTI performance.

\subsubsection{Dark Current}

\begin{figure}[htbp]
	\centering 
	\includegraphics[width=\textwidth,trim=1 0 0 0,clip]{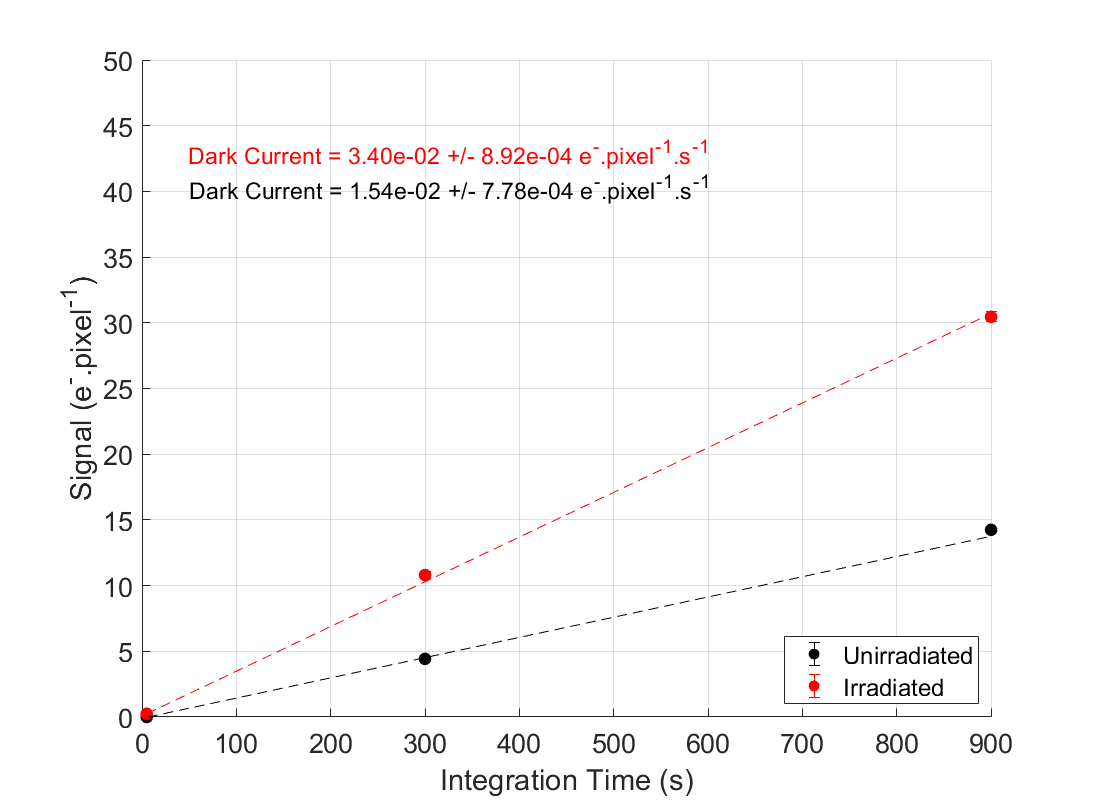}
	\caption{\label{fig:DCfit} Example dark current plot at 188~K }
\end{figure}

As dark current increases after an irradiation and has the potential to anneal the DC is measured throughout the campaign by taking images with 5, 300 and 900~s integration times to create plots such as that in Figure ~\ref{fig:DCfit}. The gradient can then be taken from the fitted line and the dark current in terms of electrons generated in a pixel per unit time can be recorded. The same red hatched area in Figure ~\ref{fig:Coldhead}b is used for the post-irradiation data. 

\section{Results}
\subsection{Trap Landscape}
\label{section:TrapResults}
In an irradiated CCD the trap population present is susceptible to temperature annealing and their concentration can change very quickly ~\cite{Moll}. It is therefore important to assess how this impacts the trap landscape by carrying out the temperature cycling described in Figure ~\ref{fig:Schedule} to give a representation of how the device is likely to respond to any temperature variations experienced throughout a space mission. 

\begin{figure}[htbp]
	\centering 
	\includegraphics[width=\textwidth,trim=1 0 0 0,clip]{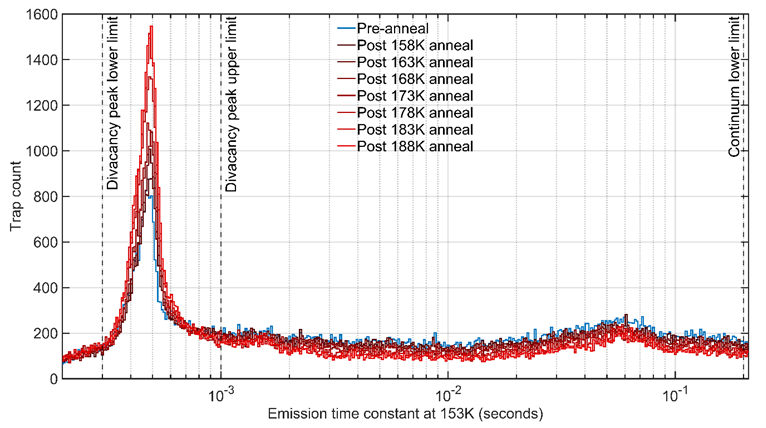}
	\caption{\label{fig:TrapLandscape} Trap landscape at 153~K after the annealing steps shown in Figure ~\ref{fig:Schedule}. Note that the limits imposed do not represent a pertinent physical boundary, just an analysis technique used by the author.}
\end{figure}

By plotting the defect landscape at 153~K after each subsequent anneal step, the evolution of the defect landscape as a function of annealing temperature can be explicitly seen (Figure ~\ref{fig:TrapLandscape}). Generally, the defect landscape is relatively stable as a function of annealing temperature, although some gradual changes are seen in different emission time constant ranges. Within the double negative state of the divacancy state peak, the number of defects gradually increases with anneal temperature, whereas the opposite occurs in the continuum to the right of the divacancy (a gradual decrease). To quantify the defect evolution, limits can once again be imposed on the emission time constant to isolate different areas independently.

\begin{figure}[htbp]
	\centering 
	\includegraphics[width=\textwidth,trim=1 0 0 0,clip]{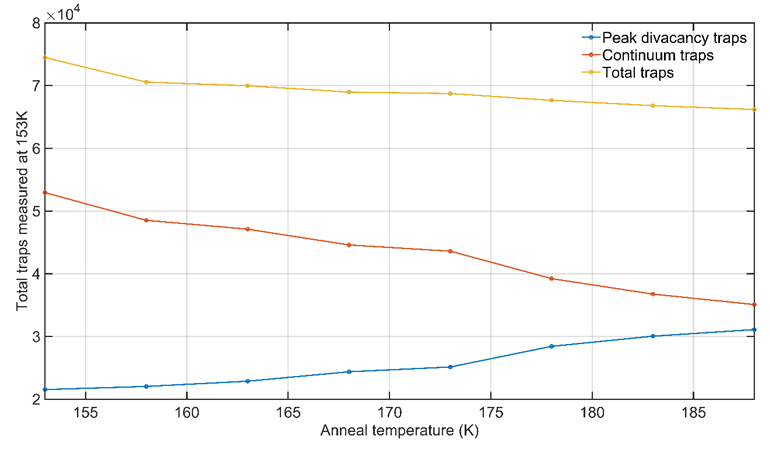}
	\caption{\label{fig:NumberofDefects} Quantification of the defects within the parameter space probed as a function of anneal temperature. Error on the total number of traps are several orders of magnitude smaller than the data points and are too small to be seen on this plot, they have been calculated by summing the errors in the fitted emission time constant in quadrature.}
\end{figure}

Three limits are seen in Figure ~\ref{fig:TrapLandscape}, two defining the peak of the divacancy, and a third defining the slower end of the continuum of defects. By plotting the quantity of defects between these limits, it is possible to see the change in quantity of defects as a function of annealing temperature and is shown in Figure ~\ref{fig:NumberofDefects}. 

Figure ~\ref{fig:NumberofDefects} shows three distinct lines; the total number of defects between the divacancy peak lower limit and the continuum lower limit (yellow lines); the number of traps in the continuum (red line), between the divacancy peak upper limit and the continuum lower limit, and the peak divacancy traps (blue), between the lower and upper limit of the divacancy, respectively. Overall, the total number of defects within the parameter space is not constant, gradually falling as a function of annealing temperature. Defects located in the continuum and divacancy peak show opposite trends as a function of annealing temperature, with a steady increase and decrease as a function of annealing temperature, respectively. At low annealing temperatures (153~K), the total number of defects are dominated by the continuum, however this is eroded as annealing temperature increases, with an almost equal number of defects in these two populations after the 188~K anneal. The number of divacancy defects located within the peak is slightly overestimated however, as the divacancy peak is situated upon a broad continuum of approximately 200 defects crossing a wide range of emission time constants, meaning that the true value is slightly lower.

Although defects within the continuum decrease, as defects within the divacancy peak increase, this does not necessarily mean that continuum defects are migrating directly into the peak of the divacancy. Further analysis could investigate the pixel locations (another output of the trap pumping technique) of both continuum and peak divacancy defects, tracking the movement of defects as a function of time. It would be clear if the pixel locations of continuum defects later contain peak divacancy defects and vice versa. For the scope of this paper however, it is clear that the defect range probed is relatively constant.

\subsection{X-ray Parallel CTI Annealing}

It is clear from the trap pumping results in Section ~\ref{section:TrapResults} that the trap landscape evolves as the anneal temperature gradually increases, however, in terms of a space mission one of the most important performance parameters that is affected by the presence of traps is the CTI. To assess this X-ray images have been taken at the same intervals as the trap pumping data and the parallel X-ray CTI measured.

\begin{figure}[htbp]
	\centering 
	\includegraphics[width=\textwidth,trim=1 0 0 0,clip]{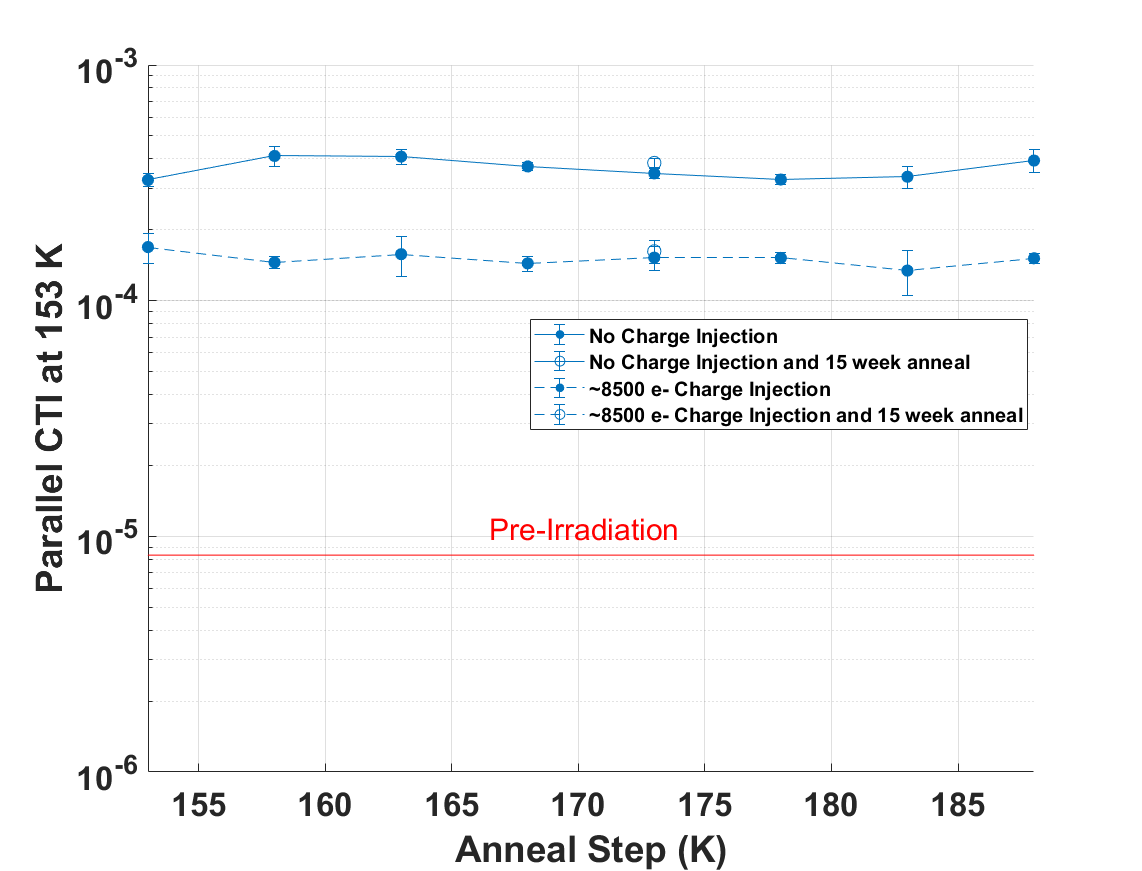}
	\caption{\label{fig:CTI_Anneal} Impact of temperature annealing on X-ray parallel CTI with and without one line of $8500$~e$ ^{-} $  charge injection. The unfilled data points on the plot show the parallel CTI after a 15~week anneal at 153~K, after warming up to 173~K. Red line indicates the pre-irradiation parallel CTI at 153~K without any charge injection }
\end{figure}

The red trace in Figure ~\ref{fig:CTI_Anneal} shows the pre-irradiation parallel CTI at 153~K and the increase in CTI and hence decrease in device performance is clear when comparing pre to post-irradiation results, the factor of degradation appears to be of the same order as previous ESA results ~\cite{ESAPLATO} from a room temperature irradiated unbiased CCD280.     

The results (solid line) in Figure ~\ref{fig:CTI_Anneal} clearly show that warming the CCD280 from 153 to 188~K before measuring again at the nominal operating temperature has no measurable impact on the parallel CTI. This remains the case with a charge injection line included (dashed line in Figure ~\ref{fig:CTI_Anneal}) and with the 15~week anneal at 153~K after warming to 173~K (empty circles in Figure ~\ref{fig:CTI_Anneal}). 

It is thought that the lack of impact annealing has on the CTI is due to the number of effective traps in the landscape only reducing slightly with a good proportion of those moving from the continuum and into the divacancy peak where their emission time constant is not different enough to change the amount they affect the CTI.

\subsection{Dark Current Annealing}

\begin{figure}[htbp]
	\centering 
	\includegraphics[width=\textwidth,trim=1 0 0 0,clip]{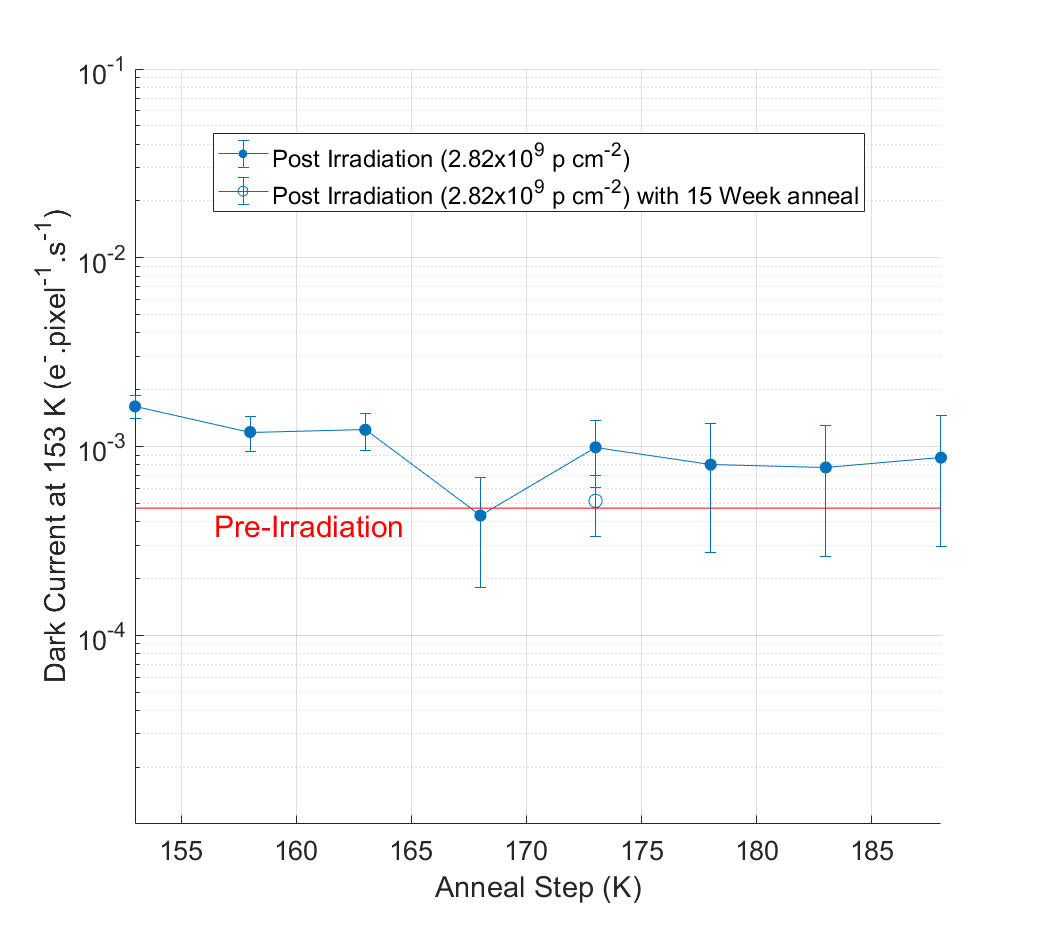}
	\caption{\label{fig:DCAnneal} Impact of temperature annealing on dark current. The unfilled data points on the plot show the dark current after a 15~week anneal at 153~K, after warming up to 173~K. Red line indicates the pre-irradiation parallel DC at 153~K}
\end{figure}

The dark current data in Figure ~\ref{fig:DCAnneal} shows very little change after the temperature and time anneal. There is potentially a slight improvement up to 173~K however it is difficult to make any solid conclusions as it is mostly within error. 

\newpage
\section{Summary}

Throughout the course of a typical earth orbit mission the focal plane detectors are damaged by proton radiation which over time causes a degradation in their performance. To understand the impact of the damage on CCDs and hence be able to prepare for countering its effects a cryogenic proton irradiation campaign has been carried out using a CCD280. 

This paper focusses on the impact of temperature cycling between 153 to 188~K. This assesses whether incrementally increasing the temperature changes the performance of the CCD due to the traps created in the irradiation annealing and no longer impacting the parallel CTI. The CTI results however show that warming up to 188~K has no detectable impact on the parallel CTI in the CCD280 and the dark current results show minimal if any improvement. Through the analysis of trap pumping data taken throughout the anneal the evolution of the trap landscape can be mapped. It is thought that whilst the trap landscape changes due to the anneal the majority of changes will not have a significant enough effect on the emission time constant of the traps to change how they degrade the parallel CTI.

In terms of using the CCD280 in a space mission the consistency of the parallel CTI across the tested temperature range shown here is a good indicator that a similar behaviour will be seen in orbit. This is beneficial since any correction of the images to counter the effect of CTI will not have to take the thermal history of the focal plane into account.

Analysis of this irradiated device performance is continuing, by investigating the operational parameter space for optimizing the read-out scheme to improve the CTI performance from the baseline reported here, and to study the device performance using frame transfer readout. 

\acknowledgments

\end{document}